\documentclass[12pt,a4paper,fleqn]{article}

\usepackage[DIV12]{typearea}
\usepackage{amsmath}
\usepackage{amssymb}
\usepackage{amsfonts}
\usepackage{amscd}
\usepackage{graphicx}
\usepackage[footnotesize]{caption2}
\usepackage{mcite}

\newcommand{\stau}{{\widetilde\tau}}
\newcommand{\snu}{{\widetilde\nu}}
\newcommand{\gravitino}{\widetilde{G}} 

\newcommand{\figref}[1]{Fig.~\ref{fig:#1}}

\newcommand{\eVdist}{\kern-0.06667em}

\newcommand{\mev}{{\,\text{Me}\eVdist\text{V\/}}}
\newcommand{\gev}{{\,\text{Ge}\eVdist\text{V\/}}}
\newcommand{\tev}{{\,\text{Te}\eVdist\text{V\/}}}

\allowdisplaybreaks[1]

\begin{document}

\begin{titlepage}
\renewcommand{\thefootnote}{\alph{footnote}}

\begin{flushright}
DESY 06-158
\end{flushright}

\vspace*{1.0cm}

\renewcommand{\thefootnote}{\fnsymbol{footnote}}

\begin{center}
{\Large\bf 
Dark Matter from Gaugino Mediation
}

\vspace*{1cm}
\renewcommand{\thefootnote}{\alph{footnote}}

\textbf{
Wilfried Buchm\"uller\footnote[1]
{Email: \texttt{wilfried.buchmueller@desy.de}},
Laura Covi\footnote[2]{Email: \texttt{laura.covi@desy.de}},
J\"orn Kersten\footnote[3]{Email: \texttt{joern.kersten@desy.de}}
\\and
Kai Schmidt-Hoberg\footnote[4]{Email: \texttt{kai.schmidt.hoberg@desy.de}}
}
\\[5mm]

Deutsches Elektronen-Synchrotron DESY, 22603 Hamburg, Germany
\end{center}

\vspace*{1cm}

\begin{abstract}
\noindent 
We study dark matter (DM) for gaugino-mediated supersymmetry breaking and compact 
dimensions of order 
the grand unification scale. Higgs fields are bulk fields, and in general 
their masses differ  
from those of squarks and sleptons at the unification scale. As a 
consequence, at different
points in parameter space, the gravitino, a neutralino or a scalar 
lepton can be the lightest 
(LSP) or next-to-lightest (NLSP) superparticle. We investigate the 
constraints from primordial 
nucleosynthesis on the different scenarios. While neutralino DM
and gravitino DM
with a $\snu$ NLSP are consistent for a wide range of parameters, 
gravitino DM with 
a $\stau$ NLSP is strongly constrained. 
Gravitino DM with a $\chi^0$ NLSP is excluded.
\end{abstract}

\end{titlepage}
\newpage

\section{Introduction}

Gaugino mediation \cite{Kaplan:1999ac,Chacko:1999mi} is an 
attractive way to introduce
supersymmetry (SUSY) breaking in higher-dimensional theories with
four-dimensional branes. 
For squarks and sleptons which are confined to these
branes this yields 
no-scale boundary conditions 
\cite{Ellis:1984bm,*Inoue:1991rk}, whereas gauginos and Higgs fields 
acquire soft SUSY
breaking masses at tree-level, since they are bulk fields.

Varying the boundary conditions for the Higgs fields at the GUT scale, 
we found in 
\cite{Buchmuller:2005ma} that, apart from the gravitino, a neutralino 
or a scalar
lepton, $\stau$ or $\snu$, can be the lightest superparticle (LSP)\@. Since a 
scalar lepton
is excluded as LSP \cite{Falk:1994es,Eidelman:2004wy}, it can only be
the next-to-lightest superparticle (NLSP)
with the gravitino as LSP, 
which is
consistent with the lower bound on the gravitino mass in gaugino mediation 
\cite{Buchmuller:2005rt}. One then obtains the $\gravitino$-$\stau$ and the
$\gravitino$-$\snu$ scenarios with $\stau$ and $\snu$ as NLSP, respectively. 

The $\gravitino$-$\stau$ scenario is particularly interesting, since it may
allow to determine the gravitino mass and spin at colliders 
\cite{Buchmuller:2004rq,*Hamaguchi:2004df,*Feng:2004yi,*Brandenburg:2005he,*Martyn:2006as,*Ellis:2006vu}.
It is well known, however, to be strongly constrained by primordial
nucleosynthesis (BBN) 
\cite{Feng:2003uy,*Fujii:2003nr,Feng:2004mt,Ellis:2003dn,*Roszkowski:2004jd}. 
In the following we therefore study the impact of such constraints
on the $\gravitino$-$\stau$
scenario in gaugino mediation where the gravitino mass is bounded from
below, leading to very long-lived $\stau$ leptons, and compare it with
the $\gravitino$-$\snu$ scenario.

The decays of a neutralino NLSP into gravitino and photon or $Z$ boson,
decaying further into hadrons, make this scenario
incompatible with BBN\@.
On the other hand, we find that a neutralino LSP as 
dominant component of dark matter is a
viable possibility in gaugino mediation.  
 
The paper is organised as follows. In Section 2 and 3 we briefly recall the
boundary conditions for gaugino mediation and the BBN constraints on 
NLSP abundances, respectively. Section 4 deals with neutralino
dark matter, and in Section 5 we discuss gravitino dark matter with
slepton NLSPs. Our results are summarised in Section 6.

\section{Superparticle Masses from Gaugino Mediation}

Consider a theory with $D$ space-time dimensions and four-dimensional branes
located at different positions in the $D-4$ compact spatial dimensions.
In models with gaugino mediation \cite{Kaplan:1999ac,Chacko:1999mi}, the gauge 
superfields live in the bulk, while the chiral superfield $S$ responsible 
for SUSY breaking
is localised on one of the four-dimensional branes.  
The Higgs fields can live in the bulk as well.

A vacuum expectation value $F_S$ for the $F$-term of $S$ breaks SUSY
and leads to 
a non-vanishing gaugino mass $m_{1/2}$ as well as a gravitino mass 
$m_{3/2}$ at the compactification scale $M_C$, which we assume to be of
order the unification scale $M_\text{GUT}$.
Like the gaugino masses, also the soft Higgs masses $m^2_{\tilde{h}_i}$ and
the parameters $\mu$ and $B\mu$ are generated from non-renormalisable
couplings with the brane field $S$. Here $\tilde{h}_1$ is the 
Higgs which couples to the down-type quarks, 
whereas $\tilde{h}_2$ is the up-type Higgs.
Neglecting small corrections to the 
scalar masses from gaugino loops  as well as corrections to the gauge 
couplings from brane-localised terms breaking the unified gauge symmetry,
one obtains the boundary conditions of gaugino mediation
with bulk Higgs fields
at the compactification scale \cite{Chacko:1999mi}:
\begin{subequations}\label{gauginomed}
\vspace*{-3ex}
\begin{align}
&g_1 = g_2 = g_3 = g \simeq 1/\sqrt{2} \;,
\\
  &M_1 = M_2 = M_3 = m_{1/2} \;,
\\
  &m_{\tilde \phi_\mathrm{L}}^2 =
  m_{\tilde \phi_\mathrm{R}}^2 = 0
  \quad \text{for all squarks and sleptons } \tilde\phi \;,
\\
  &A_{\tilde\phi} = 0
  \quad \text{for all squarks and sleptons } \tilde\phi\;,
\\
  &\mu, B\mu, m^2_{\tilde{h}_i} \neq 0 \quad (i=1,2) \;,
\end{align}
\end{subequations}
where the GUT charge normalisation is used for $g_1$.
If the Higgs fields are localised on a brane, one has
$m^2_{\tilde{h}_i}=B\mu=0$, which is a special case of minimal supergravity.

The ranges of SUSY breaking parameters leading to a viable
low-energy spectrum have been discussed in \cite{Buchmuller:2005ma}. 
The spectrum is determined by the boundary conditions
(\ref{gauginomed}) and the renormalisation group equations. 
The model favours
moderate values of $\tan\beta$ between about 10 and 25. Much smaller and larger
values are in conflict with the LEP lower bounds on the Higgs mass and 
the $\stau$ mass, respectively. The gaugino
mass at the GUT scale cannot be far below $500\gev$ in order to
satisfy the LEP bound on the Higgs mass. 
Typically, the lightest
neutralino is bino-like with a mass of $200\gev$, and the gluino mass is
about $1.2\tev$.  
Depending on $m^2_{\tilde{h}_1}$, 
either the right-handed or the left-handed sleptons
can be lighter than the neutralinos.  The corresponding region in
parameter space grows with $\tan\beta$.  

Gaugino mediation gives a lower bound on the gravitino mass.
This bound depends on $m_{1/2}$, the number of space-time dimensions 
and the compactification scale \cite{Buchmuller:2005rt}.
Motivated by a six-dimensional orbifold GUT \cite{Asaka:2003iy},
we choose $D=6$ and $M_C=M_\text{GUT}$ leading to
$m_{3/2} \gtrsim 0.1 \cdot m_{1/2} \gtrsim 50\gev$.
As the lower limit on $m_{3/2}$ was derived 
using na\"ive dimensional analysis  \cite{Chacko:1999hg},
it can well be relaxed by a factor of order one.
We therefore also consider $m_{3/2}=10 \gev$ as a conservative lower
bound.  Note that varying $D$ between $5$ and $10$, the lower bound
ranges between $20\gev$ and $0.1\gev$.

\section{BBN Constraints on the Abundance of NLSPs}

Big Bang Nucleosynthesis starts about $100\,\text{s}$ after the
big bang at a temperature of about $0.1\mev$.  In the scenario where
the gravitino is the LSP, the NLSP decays considerably later.
The decay products of such long-lived particles can alter the 
primordial light element
abundances \cite{Cyburt:2002uv,Kawasaki:2004qu,Jedamzik:2006xz}.
This leads to constraints on the released
electromagnetic and hadronic energy. 
To a good approximation these constraints can be quantified by 
upper bounds on the product  $\epsilon_\text{em,had} Y_\text{NLSP}$.
Here $\epsilon_\text{em,had}$ is the average electromagnetic or
hadronic energy emitted in a single NLSP decay
and the abundance $Y_{\text{NLSP}}$ is given by the NLSP number density
prior to decay divided by the total entropy density,
\begin{equation} \label{eq:YNLSP}
	Y_{\text{NLSP}} \equiv \frac{n_\text{NLSP}}{s} \;.
\end{equation}
We determine this normalised NLSP number density numerically, 
assuming that the NLSP freezes
out with its thermal relic density.

For our analysis we use the bounds compiled in Fig.~9 of
\cite{Steffen:2006hw} (see also Fig.~\ref{fig:tauYtb10} below), 
which were computed in the earlier studies
\cite{Cyburt:2002uv,Kawasaki:2004qu}. 
These bounds assume an NLSP
mass of $1\tev$, but since they are quite insensitive to this
mass, we will use them here, too. 
Furthermore, they assume that there 
is no entropy production between the decoupling of the NLSP and the
start of BBN\@.
As there is still considerable uncertainty in the
measurements of the primordial element abundances, \cite{Steffen:2006hw}
used two different data sets, giving ``severe'' and ``conservative''
limits.  The severe limits are derived from
\begin{subequations} \label{eq:SevereBounds}
\begin{align}
	2.02 \cdot 10^{-5} &< \frac{n_\text{D}}{n_\text{H}} <
	3.66 \cdot 10^{-5} \;,
\\
	0.227 &< Y_p < 0.249 \;,
\end{align}
\end{subequations}
where $n_\text{X}$ is a primordial number density and $Y_p$ the
primordial mass fraction of $^4$He.  
The relevant conservative limits are derived from
\begin{subequations} \label{eq:ConservativeBounds}
\begin{align}
	1.3 \cdot 10^{-5} &< \frac{n_\text{D}}{n_\text{H}} <
	5.3 \cdot 10^{-5}  \;,
	\label{eq:DConservativeEm}
\\
	0.231 &< Y_p < 0.253 \;.
\end{align}
\end{subequations}
Note that although the upper bounds on the Deuterium abundance leading to the
conservative and severe constraints differ by less than a factor of two,
the resulting bounds on $\epsilon_\text{em} Y_\text{NLSP}$ differ by an
order of magnitude.

The constraints on hadronic and electromagnetic energy release are
assumed to be independent, although there can be cancellations between
them in special cases. 
We consider points in parameter space violating the conservative limits
to be ``excluded'', but points violating only the severe limits to
be ``disfavoured''.
The observed abundances of $^3$He, $^6$Li and $^7$Li are
not used, since they still suffer from large systematic uncertainties.

\section{Neutralino Dark Matter}

We are now in a position to determine the 
cosmologically allowed, disfavoured and excluded 
regions of the parameter space
of models with gaugino-mediated supersymmetry breaking.
Since moderate values of $\tan\beta$ are favoured,
we consider the cases $\tan\beta=10$ and $\tan\beta=20$ in the following.
As a benchmark point for our discussion we take
the unified gaugino mass to be $m_{1/2}=500\gev$ and the supersymmetric
Higgs mass parameter to be positive, sign $\mu >0$.

Both for the constraints from BBN and for those from the observed
cold dark matter density, the abundance $Y_{\text{(N)LSP}}$ 
of the (N)LSP is essential.
In this section we will discuss
neutralino dark matter and then turn to gravitino dark matter with slepton
NLSPs in the next section.

\subsection{Calculation of the Abundance}

We use micrOMEGAs 1.3.6 \cite{Belanger:2001fz,*Belanger:2004yn} to
calculate the abundance and the energy density of the (N)LSP
numerically.  The superpartner spectrum is determined by 
SOFTSUSY 2.0.6 \cite{Allanach:2001kg}.
For the top quark pole mass, we use the latest best-fit value of
$172.5\gev$ \cite{Tevatron:2006qt}.\footnote{In addition, we use
$m_b(m_b)=4.25\gev$ and 
$\alpha_s^{\text{SM }\overline{\text{MS}}}(M_Z)=0.1187$, the default
values of SOFTSUSY.  Some other SM parameters are hard-coded in
micrOMEGAs, 
$\alpha_\text{em}^{-1 \text{ SM }\overline{\text{MS}}}(M_Z)=127.90896$,
$G_F = 1.16637 \cdot 10^{-5} \gev^{-2}$, and
$m_\tau = 1.777\gev$.}

We first consider the case where a neutralino is lighter than all
sleptons and squarks, so that it is an LSP or NLSP candidate.  In the
corresponding parameter space region for $\tan\beta=20$, we find
numerically
\begin{subequations}
\begin{align}
	2.6 \cdot 10^{-13} &\leq Y_\chi \leq 5.0 \cdot 10^{-12} \;,
\\
	83.0\gev &\leq m_\chi \leq 204\gev \;,
\\
	8.15 \cdot 10^{-3} &\leq \Omega_\chi h^2 \leq 0.273 \;.
\end{align}
\end{subequations}
For $\tan\beta=10$, the results are very similar, with a slightly
larger maximal abundance of $8.7 \cdot 10^{-12}$.
Here the relation between neutralino relic abundance $Y_{\chi}$
and energy density $\Omega_\chi h^2$
is given by
\begin{equation}
Y_{\chi} = \frac{\Omega_\chi \rho_c}{s \, m_{\chi}} \simeq 3.64\cdot 10^{-11}
           \left(\frac{100\gev}{m_\chi}\right)\Omega_\chi h^2 \;,
\end{equation}
with $\rho_c$ the critical density and $s$ the entropy density of the universe.

\subsection{Neutralino LSP}

Gaugino mediation provides only a lower bound on the mass of the
gravitino.  Therefore, it may well be quite heavy, and the lightest
neutralino may be the LSP.  In this case, decays of the long-lived
gravitino threaten the success of BBN, which leads to an upper bound on
the gravitino density and thus on the reheating temperature
\cite{Cyburt:2002uv,Kawasaki:2004qu}.  
The other superparticles decay into the LSP
before the start of BBN and do not cause problems, unless LSP and NLSP
are nearly degenerate.  For example, if the NLSP is a stau, BBN
constraints become potentially important for 
$m_\stau-m_\chi \lesssim 100 \mev$ \cite{Jittoh:2005pq}.  We neglect
this possibility, since the corresponding region in the parameter space
is tiny.

\begin{figure}
\centering
\includegraphics{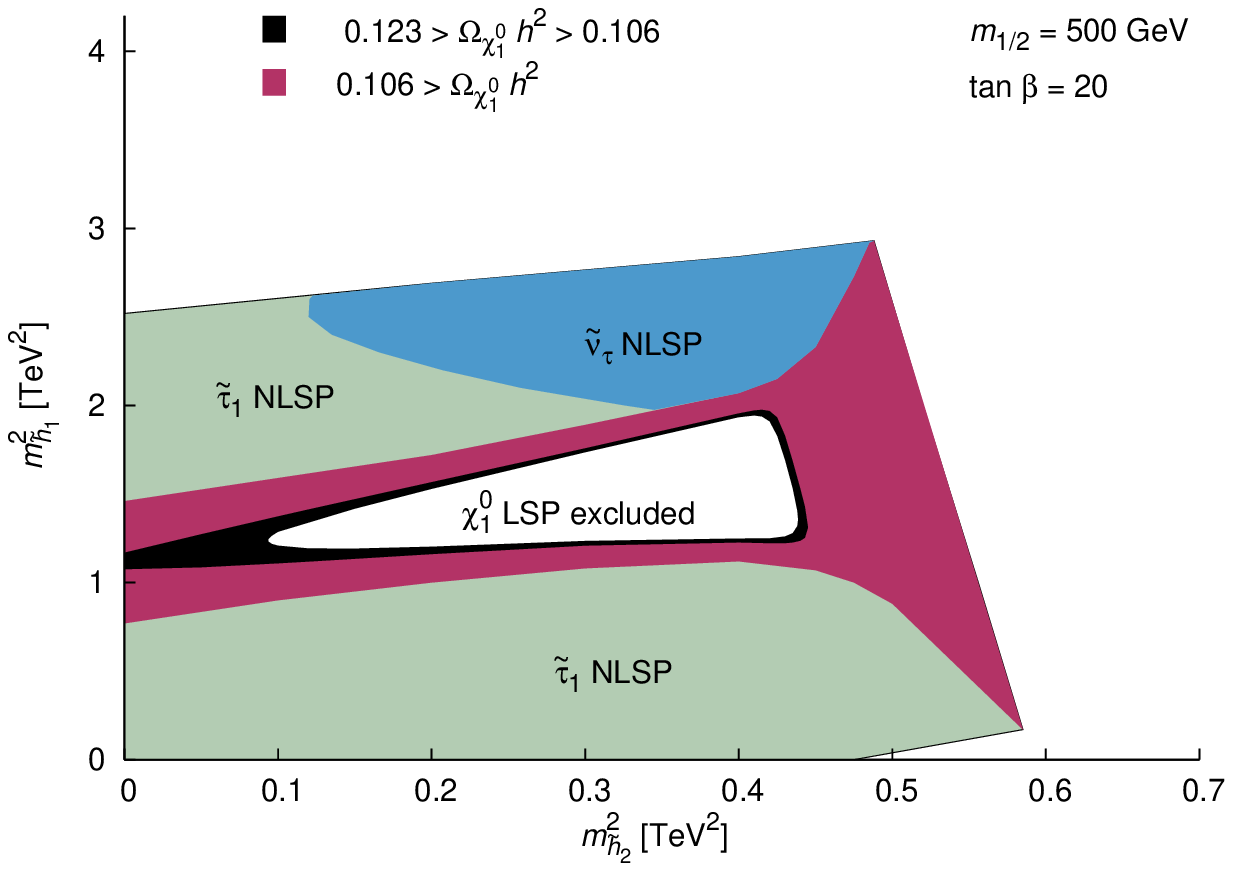}\\[2mm]
\includegraphics{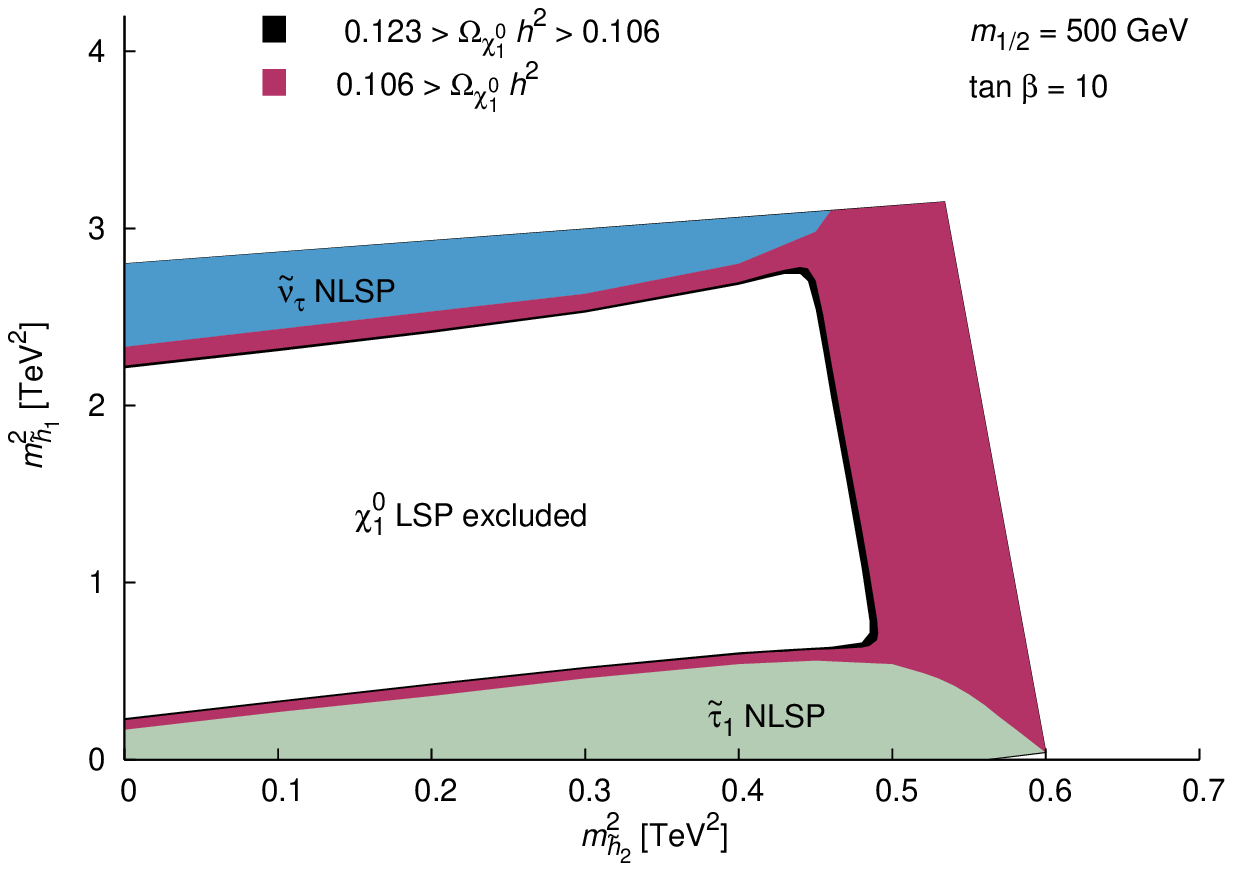}
\caption{Allowed region for the soft Higgs masses for $m_{1/2}=500\gev$
and $\tan\beta=20$ ($\tan\beta=10$). A neutralino is lighter 
than all sleptons in the white, black and magenta (dark-gray) area.
The upper limit on $\Omega_\chi h^2$ excludes the white region,
whereas in the magenta (dark-gray) area  $\Omega_\chi h^2$ is smaller
than the observed cold dark matter density. 
The correct
 dark matter density is obtained in the black region.  
In the green (light-gray) 
and blue (medium-gray) areas a 
 slepton is the NLSP.
}
\label{fig:snu_parameterspace}
\end{figure}

This leaves the observed dark matter density as the only constraint on
the neutralino LSP scenario we have to consider.  We use the $3\sigma$
range given in \cite{Seljak:2006bg}%
\footnote{The analysis (labelled ``All Data $-$ LYA'') used the 
 measurements of the CMB power spectrum (temperature and polarisation)
 by WMAP (3-year data) and other experiments, the SDSS and 2dF galaxy
 clustering analyses, the SDSS luminous red galaxy constraints on the
 acoustic peak, as well as the Gold and the SNLS supernovae samples.
},
\begin{equation} \label{eq:OmegaDMObs}
    0.106 < \Omega_\text{DM} h^2 < 0.123 \;.
\end{equation}
The upper limit excludes the white regions in
\figref{snu_parameterspace}.  Since the dark
matter could be made up of several components and since non-thermal
production could be significant, we have two viable regions in parameter
space.  In the first one, the thermal neutralino relic density falls
into the range \eqref{eq:OmegaDMObs} and hence this particle makes
up all the dark matter. This region is shown in black in
\figref{snu_parameterspace}.  There the bino
contributes at least $75\%$ ($80\%$) to the lightest neutralino for
$\tan\beta=10$ ($\tan\beta=20$).
The missing $25\%$ ($20\%$) come from the two Higgsinos, while the wino component
of $\sim 1\%$ is negligible.
On the left edge the lightest neutralino is a pure bino.
The second viable region is shown in magenta (dark-gray) in the figure.
Here the thermal
neutralino density is smaller than the lower bound in 
Eq.~\eqref{eq:OmegaDMObs} and hence only
constitutes a part of the dark matter density.
The lightest neutralino is almost a pure Higgsino
at the right edge of the parameter space. 

In most parts of the parameter space, some tuning is necessary if
neutralinos are to make up all the dark matter.  This is very similar to
what has been found in other scenarios for SUSY breaking, for instance
in mSUGRA (see for example~\cite{Gomez:2004ek,*Ellis:2004tc,*Djouadi:2006be}). In part, the reason is simply
that the dark matter density has been measured rather accurately.
For $\tan\beta=20$ and small $m^2_{\tilde{h}_2}$, the situation looks somewhat
better.  We are not aware of a simple physical explanation for this.
Apparently, for $m^2_{\tilde{h}_2} < 0.1\tev^2$ the maximum of
$\Omega_\chi$ as a function of $m^2_{\tilde{h}_1}$ lies in the
experimentally allowed region.  Around the maximum, a change in
$m^2_{\tilde{h}_1}$ leads only to a relatively small change in
$\Omega_\chi$, so that the energy density remains in the favoured
range in a rather broad strip of parameter space.  For larger
$m^2_{\tilde{h}_2}$, the maximum value of $\Omega_\chi$ is too large.
Consequently, it depends rather sensitively on $m^2_{\tilde{h}_1}$ in
the allowed region, and thus this region is narrow.

Let us finally comment on the direct detection of neutralino dark matter 
in our scenario. As in the general MSSM case, the detection 
cross-section is suppressed for a pure bino, since the Higgs and
$Z$ exchange require a Higgsino component. In fact, for 
$\tan\beta = 10$, $m^2_{\tilde{h}_1}= 2.21\tev^2$ and 
$m^2_{\tilde{h}_2} = 0 $, one obtains 
$\sigma_{\chi p,n} = 9 \cdot 10^{-13} \,\text{nb}$ for the spin-independent 
cross-section per nucleon~\cite{fornengo},
whereas the present bound on this cross-section
is of the order of $10^{-9}\,\text{nb}$~\cite{Akerib:2005kh}.
The cross-section is larger in the region with a larger Higgsino
component, where for $\tan\beta = 10$, 
$m^2_{\tilde{h}_1}= 2.76 \tev^2$, 
$m^2_{\tilde{h}_2} = 0.44 \tev^2$, one obtains  
$\sigma_{\chi p,n} = 4 \cdot 10^{-11} \,\text{nb}$ \cite{fornengo}.
Although the cross-section is at least one 
order of magnitude below the present bounds, 
it could be reached by the next generation of dark matter experiments 
\cite{Bauer:2005xk}.

\subsection{Neutralino NLSP}

With a light gravitino, a scenario with a gravitino LSP and a neutralino
NLSP is possible, too.  However, it turns out that this is ruled out by
the BBN constraints in gaugino mediation.

The region where $m_\chi > m_{3/2} + m_Z$ is certainly excluded by the
hadronic BBN constraints for all gravitino masses we consider, since the
two-body decay $\chi^0_1 \to Z \gravitino$ is possible and since
the hadronic branching ratio of the $Z$ is large
\cite{Feng:2004mt,Cerdeno:2005eu}.  However, the situation
is less clear for lighter
neutralinos when the two-body decay into real $Z$ bosons is not
possible.  For $m_{3/2}=50\gev$, this is the case for $m_\chi<141\gev$.
The corresponding parameter space region lies at the right end of the
allowed region, where $m^2_{\tilde{h}_2} \gtrsim 0.5\tev^2$.  In this
region, the $\mu$ parameter is rather small \cite{Buchmuller:2005ma}, so
that there is significant mixing between the neutralinos.  The Higgsino
components of the lightest neutralino lead to a relatively large
annihilation cross section and thus to a relatively small abundance.
From Fig.~9 of \cite{Steffen:2006hw} we can read off that the 
severe hadronic bound is never stronger than
\begin{equation}
\label{eq:hadbound}
\epsilon_{\text{had}}Y_{\text{NLSP}} \lesssim 5\cdot 10^{-14} \gev
\end{equation}
for any NLSP lifetime.
With the estimate 
$\epsilon_\text{had}^\chi\simeq\frac{2}{3}(m_\chi-m_{3/2})\cdot 10^{-3}$
from \cite{Feng:2004mt}, we find
\begin{equation}
	1.0 \cdot 10^{-14} \gev \leq \epsilon_\text{had}^\chi Y_\chi \leq
	2.1 \cdot 10^{-14} \gev \;,
\end{equation}
which is well below the stringent hadronic bound (\ref{eq:hadbound}).
Considering
electromagnetic energy release instead, we have
\begin{equation}
	\epsilon_\text{em}^\chi \simeq \frac{m_\chi^2-m_{3/2}^2}{2m_\chi}
\end{equation}
and
\begin{equation}
	\Gamma_\chi = 
	\frac{|N_{11} \cos\theta_W + N_{12} \sin\theta_W|^2 \, m_\chi^5}
	 {48\pi \, m_{3/2}^2 M_\text{P}^2} \,
	\biggl( 1-\frac{m_{3/2}^2}{m_\chi^2} \biggr)^3 \,
	\biggl( 1+3\frac{m_{3/2}^2}{m_\chi^2} \biggr) 
\end{equation}
for the width of the dominant neutralino decay mode $\chi^0_1 \to \gamma
\gravitino$,
where $N_{1i}$ are elements of the neutralino mixing matrix, so that
e.g.\ $|N_{11}|^2$ is the bino fraction \cite{Feng:2004mt}.
This leads to
\begin{align}
	1.1 \cdot 10^{-11} \gev &\leq \epsilon_\text{em}^\chi Y_\chi \leq
	2.2 \cdot 10^{-11} \gev \;,
\\
	3.3 \cdot 10^8\,\text{s} &\leq \tau_\chi \leq 
	4.1 \cdot 10^{10}\,\text{s}
\end{align}
for both $\tan\beta=10$ and $\tan\beta=20$.
Comparing with the electromagnetic limits in \figref{tauYtb10}, we
see that even the conservative BBN bound is violated.  This result
remains true for $m_{3/2}=10\gev$.
Thus, we conclude that a neutralino NLSP with a mass below $m_Z+m_{3/2}$
is excluded by the BBN constraints on electromagnetic energy
release.  Consequently, the lightest neutralino is not a viable NLSP
candidate in gaugino mediation.

\section{Gravitino Dark Matter with Slepton NLSPs}

\subsection{Lifetime of Slepton NLSPs}
The slepton decay rate is dominated by the two-body decay into lepton
and gravitino,
\begin{equation}\label{eq:2bodydecay}
 \Gamma_{\tilde{l}}^\text{2-body} =
 \frac{m_{\tilde{l}}^5}{48\pi\,m_{3/2}^2\,M_\mathrm{P}^2} \,
	\biggl(1-\frac{m_{3/2}^2}{m_{\tilde{l}}^2}\biggr)^4 \;,
\end{equation}
where $m_{\tilde{l}}$ is
the slepton mass, $M_\text{P} = 2.4 \cdot 10^{18} \gev$ is the reduced
Planck mass, and where the lepton mass has been neglected.
With a typical largest slepton mass of around 
$200 \gev$ in the $\tilde{l}$ NLSP region
and the smallest gravitino mass of $10\gev$ this leads to 
a lower bound on the slepton lifetime of
\begin{equation} \label{eq:StauLifetime}
\tau_{\tilde{l}} \gtrsim 1.8 \cdot 10^5 \,\text{s}\;,
\end{equation}
which is a time where the BBN constraints become stringent.

\subsection{Stau NLSP}

For both values of $\tan\beta$, imposing the lower bound from collider
searches \cite{Eidelman:2004wy}, we find
\begin{equation}
	86\gev < m_\stau \leq 203\gev
\end{equation}
in the $\stau$ NLSP region. The upper limit on the stau mass 
within this region only depends on the mass of the lightest neutralino
and is therefore almost
independent of $\tan\beta$. With $m_{3/2}=50\gev$, this mass range corresponds 
to the range
\begin{equation}
	5.5 \cdot 10^6 \,\text{s} \leq \tau_\stau \leq 1.6 \cdot 10^9 \,\text{s}
\end{equation}
for the lifetime.  If we restrict ourselves to stau masses above
$100\gev$, the upper bound is lowered to $4.6 \cdot 10^8 \,\text{s}$.

\begin{figure}
\centering
\includegraphics{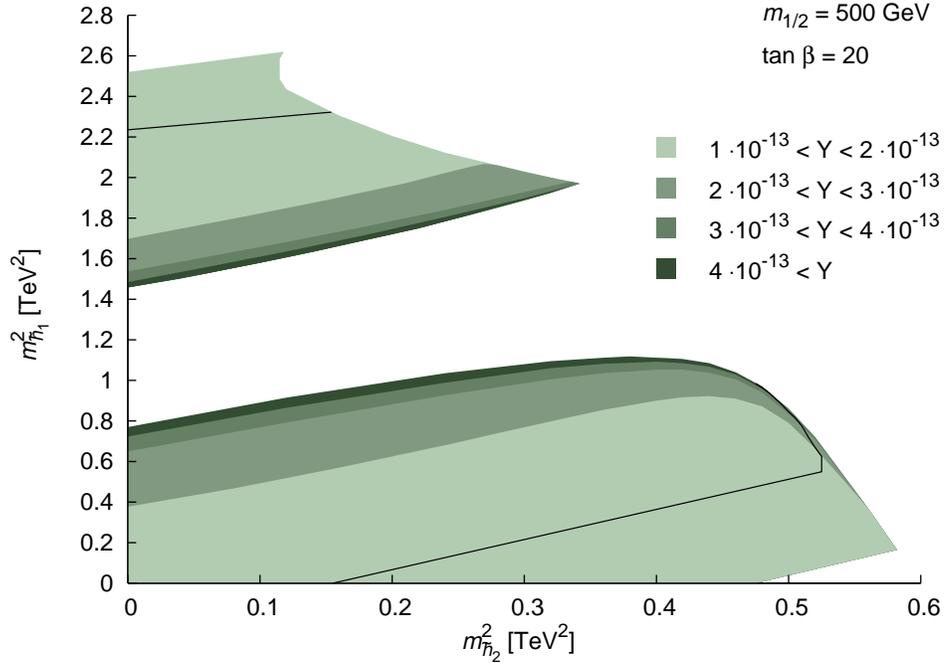}
\caption{Stau abundance obtained numerically with micrOMEGAs in the
$\stau$ NLSP region.
The area below (lower region) and above (upper region) the black line 
is excluded by the conservative electromagnetic BBN constraints
for $m_{3/2}=50 \gev$ and $\epsilon_\text{em}=0.5 E_\tau$.}
\label{fig:YStautb20}
\end{figure}

The stau abundance in the $\gravitino$-$\stau$ scenario for
$\tan\beta=20$ is shown in \figref{YStautb20}.
We find
\begin{align}
	1.3 \cdot 10^{-13} &\leq Y_\stau \leq 6.2 \cdot 10^{-13} \;.
\end{align}
The abundance is smallest in those parts of the parameter space
where the lightest stau masses are reached.  These are the lower right
corner of the bottom region and the upper border of the upper region.
Conversely, we find the largest values close to the neutralino NLSP
region, where $m_\stau$ is largest.  Both qualitatively and
quantitatively, the situation is very similar for $\tan\beta=10$, except
that in this case the top $\stau$ NLSP region does not exist.
The approximation
\begin{equation} \label{eq:ApproxY}
	Y_\stau \simeq 
	1.2 \cdot 10^{-13} \left( \frac{m_\stau}{100\gev} \right)
\end{equation}
gives the stau abundance with
 a relative error of less than $10\%$ for the largest part ($>80\%$)
of the parameter space, where coannihilation is less important
(i.e.\ the part not too close to the neutralino LSP region).

The average hadronic energy release $\epsilon_\text{had}$ 
from a stau
decay is smaller than $10^{-2}\gev$ for stau masses around $200\gev$ 
\cite{Steffen:2006hw}. 
Consequently, in our case the combination
$\epsilon_\text{had} Y_\stau$ never exceeds $10^{-14}\gev$, so that 
even the stringent hadronic BBN bound (\ref{eq:hadbound}) is always satisfied.

However, the electromagnetic bounds are significantly more constraining.
Using
\begin{equation}
	E_\tau = \frac{m_\stau^2-m_{3/2}^2+m_\tau^2}{2m_\stau}
\end{equation}
for the energy of the $\tau$ produced in the dominant two-body $\stau$
decay and
\begin{equation}
	\epsilon_\text{em} = x E_\tau
\end{equation}
for the electromagnetic energy release, we find ($\tan \beta = 20$)
\begin{equation}
	x \cdot 3.7 \cdot 10^{-12}\gev \leq \epsilon_\text{em} Y_\stau \leq
	x \cdot 5.9 \cdot 10^{-11}\gev \;.
\end{equation}
The results for $\tan\beta=10$ fall into the same range, but with a
slightly smaller spread.
Here a part of the $\tau$ energy is lost to neutrinos
and it is not exactly known which fraction $0.3 \leq x \leq 1$ 
of the energy ends up in an electromagnetic shower.

\begin{figure}
\centering
\includegraphics{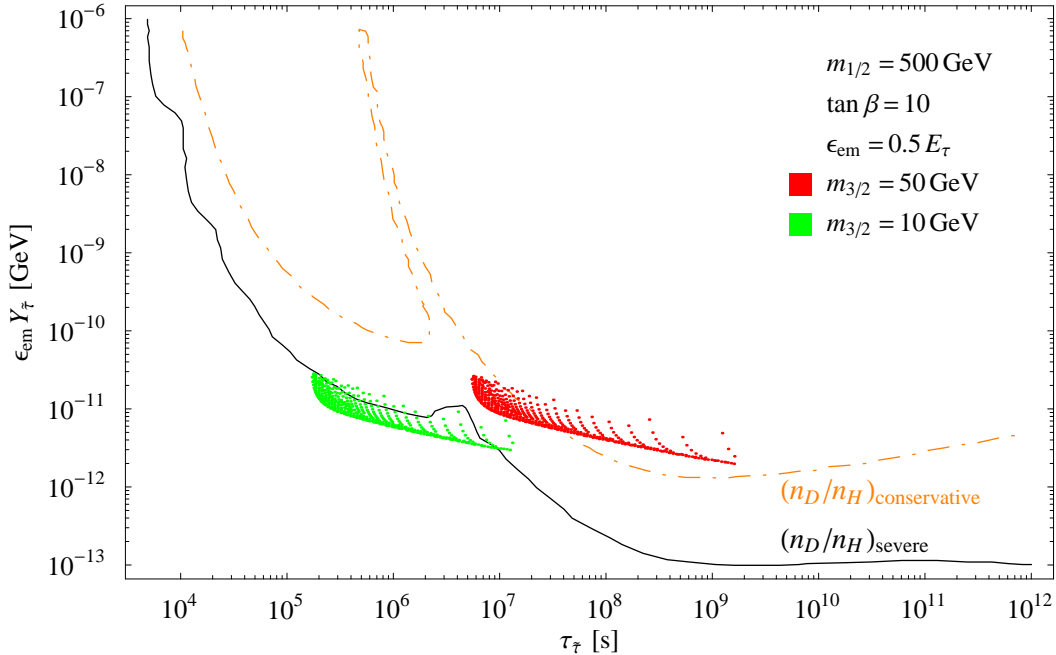}
\caption{Points from the $\stau$ NLSP region ($\tan\beta=10$) in the
 $\epsilon_\text{em} Y_\stau$\,--\,$\tau_\stau$ plane for $x=0.5$ and
 $m_{1/2}=500\gev$, obtained by scanning over $m^2_{\tilde{h}_1}$ and
 $m^2_{\tilde{h}_2}$ with a step size of $5 \cdot 10^{-3} \tev^2$ in
 both parameters.  We
 show the results for two values of the gravitino mass,
 $m_{3/2}=50\gev$ (red or dark-gray) and
 $m_{3/2}=10\gev$ (green or light-gray).
 The solid black and dash-dotted orange lines show the severe and
 conservative electromagnetic
 BBN constraints from Fig.~9 of \protect\cite{Steffen:2006hw}.
 (Only the constraints derived from the deuterium abundance are shown,
 since those from the $^4$He abundance are not relevant in the
 $\stau$ NLSP region.)  
}
\label{fig:tauYtb10}
\end{figure}

In \figref{tauYtb10}, we plot points from the $\stau$ NLSP region
for $\tan\beta=10$ in the $\epsilon_\text{em} Y_\stau$\,--\,$\tau_\stau$
plane, assuming $x=0.5$.  The red (dark-gray) points are the results 
for a gravitino mass of $50\gev$.  We also show the severe
and conservative BBN constraints from the lower plot in Fig.~9 of
\cite{Steffen:2006hw}.  We find that the severe constraints are always
violated, while the conservative constraints can be satisfied.
The same can be seen from \figref{YStautb20} for the case of
$\tan\beta=20$, where the solid black lines mark the boundary between
the parameter space regions allowed and excluded by the conservative BBN
bounds.  In the lower region, the area below and to the right of the
line is excluded.  In the upper region, this is the case for the area
above the black line.  Thus, it turns out that actually the largest part
of the parameter space is allowed.  The remaining part is typically
excluded not because of an unusually large stau abundance but because of
a too long lifetime due to a relatively small stau mass.

The severe BBN bounds can be satisfied, if the NLSP lifetime is shorter.
This is the case for smaller gravitino masses.  For $m_{3/2}=10\gev$, we
see from the green (light-gray) points in \figref{tauYtb10} that large
parts of the stau NLSP region are allowed by the severe constraints.
The conservative constraints are always satisfied in this example.

Increasing the unified gaugino mass $m_{1/2}$ leads essentially to a
rescaling of the superparticle spectrum.  Since the NLSPs become
heavier, their yield is larger.  On the other hand, the lifetime
decreases significantly, since it depends on $m_{\tilde l}^5/m_{3/2}^2$.
As a consequence, a larger part of the $\stau$ NLSP region is compatible
with the electromagnetic BBN constraints.  The hadronic constraints are
still easily satisfied unless the stau mass is close to a TeV.

Recently it has been argued that metastable charged particles alter 
BBN via the formation of bound states
\cite{Pospelov:2006sc,*Kohri:2006cn,*Kaplinghat:2006qr}.
This could lead to significantly 
more restrictive constraints on the allowed relic 
abundance of these charged particles than those we considered here.
In \cite{Cyburt:2006uv} bound-state effects in the CMSSM and mSUGRA
are studied.  The conclusion is reached
that $\stau$ NLSPs with lifetimes longer than
$10^3$\,--\,$10^4\,\text{s}$ are excluded.  If it turns out that this
statement also holds in a more general framework than the CMSSM, the
$\gravitino$-$\stau$ scenario will be ruled out for 
$m_{3/2} \gtrsim 10\gev$
unless there is sizeable entropy production between the decoupling
of the staus from the thermal bath and the start of BBN.

\subsection{Sneutrino NLSP}
The region of sneutrino NLSP corresponds to large $m^2_{\tilde{h}_1} $
and the sneutrino masses and lifetimes lie roughly in the same range 
as those of the staus, but with a somewhat larger minimal mass.
The relic abundance is in the narrow window 
\begin{equation}
\label{eq:snuabundance}
1.3 \cdot 10^{-13} \, \leq Y_{\tilde \nu} 
\leq 4.6 \cdot 10^{-13}
\,\; .
\end{equation}

The BBN bounds on a sneutrino NLSP are rather weak,
since the neutrinos emitted in the dominant two-body decay 
$\tilde{\nu} \rightarrow \nu \gravitino$ interact much less
than charged particles 
with the light nuclei. 
Nevertheless the very energetic anti-neutrinos (neutrinos) produced in such 
a decay can annihilate with the background neutrinos (anti-neutrinos)
and give rise to $e^{+} e^{-} $ pairs which contribute to 
electromagnetic showers.\footnote{
We expect the chemical potential of the sneutrinos to be negligible,
since the lepton number asymmetry can be quickly transferred into
the light leptons by the scatterings 
$\tilde\nu\tilde\nu \rightarrow \nu\nu, \ell\ell$. Therefore we
have an equal number of sneutrinos and anti-sneutrinos, so that
the NLSP decays produce both neutrinos and anti-neutrinos.
} 
Furthermore, there are contributions from
$\tilde{\nu} \rightarrow \nu \gravitino \ell \bar \ell$, but such decays
have a very small branching ratio giving
$\epsilon_{\text{em}}$ safely below $ 0.1 \gev $ and are therefore 
negligible even for our maximal value of $Y_{\tilde \nu}$.

The effects of highly energetic (anti-)neutrinos on BBN
have been studied in \cite{Gratsias:1990tr,deLaix:1993gr}
and \cite{Kawasaki:1994bs}, 
but in the last reference only for the specific
case of an unstable gravitino decaying into sneutrino and neutrino. 
Assuming $(n_\text{D}+n_\text{$^3$He})/n_\text{H} 
\lesssim 4 \cdot 10^{-5}$ 
gives $Y_X \lesssim 4\cdot 10^{-12}$ \cite{deLaix:1993gr} for a general 
relic with mass around $200\gev$ and lifetime $10^7\,\text{s}$ 
decaying equally into neutrinos and anti-neutrinos. 
Shorter lifetimes are not discussed in this work, but according 
to \cite{Gratsias:1990tr} all constraints disappear for lifetimes 
shorter than $10^6\,\text{s}$, since at those earlier times 
high-energy photons thermalise efficiently scattering off the 
CMB before having the chance to interact with the light nuclei.
Also the limits relax for longer lifetimes, since the density of
background neutrinos becomes more diluted.
In \cite{Kawasaki:1994bs} instead, the upper bound 
$(n_\text{D}+n_\text{$^3$He})/n_\text{H} 
\leq 10^{-4} $ is used and the constraints 
are given only in the $T_\text{R}$\,--\,$m_{3/2}$ plane. 
We can rephrase the strongest bound
on $T_\text{R}$ in terms of $Y_{3/2}$ giving
\begin{equation}
\label{eq:Klimit}
Y_{3/2} \lesssim 10^{-12} \quad
\mbox{for}\quad \tau_{3/2} \sim 10^7 \,\text{s}\;.
\end{equation}
Again the bound becomes quickly much weaker for longer 
or shorter gravitino lifetimes.

The maximal abundance (\ref{eq:snuabundance}) is an order of magnitude
below the limit of \cite{deLaix:1993gr}.
We cannot directly apply the limit (\ref{eq:Klimit}) 
of  \cite{Kawasaki:1994bs} to
our case due to the different dependence on lifetime and mass 
in the sneutrino case, but we note that if we take the maximal bound
\eqref{eq:Klimit} and rescale it to match the ``severe'' value 
$(n_\text{D}+n_\text{$^3$He})/ n_\text{H} \leq 3.66 \cdot 10^{-5}$, we
get an upper bound of 
$Y_{\tilde\nu}  \leq 3.66 \cdot 10^{-13} $, which is slightly
smaller than our maximal abundance.\footnote{
On general grounds the amount of D$+^3$He overproduced
should be proportional to the decaying particle abundance
$Y_X$, so we can simply rescale the constraint on $Y_X$ 
by the factor that brings their D$+^3$He abundance down
to our ``severe'' value $3.66 \cdot 10^{-5}$.
We are then imposing the bounds on Deuterium alone and
not on the sum  D$+^3$He, which provides us with a more
stringent constraint.}
So electromagnetic showers in the $\gravitino$-$\snu$ scenario surely 
do not violate the conservative constraints, but the severe 
limits might become relevant for short sneutrino lifetimes.
A more detailed analysis
appears appropriate to draw more definite conclusions.

Let us now turn to the hadronic constraints.
As for the case of the $\tilde \tau $ NLSP, the radiative
decay producing $q \bar q $ pairs via an intermediate gauge
boson has only a small branching ratio. 
The branching ratio for this decay has not 
yet been explicitly computed, 
but we can estimate it from the $\tilde\tau$ result.
In fact the diagrams mediating
the decay $\tilde\nu_\text{L} \rightarrow \nu \gravitino q \bar q$ 
are topologically the same as those involved in 
$\tilde\tau_\text{R} \rightarrow \tau \gravitino q \bar q$, apart 
from the fact that the intermediate gauge bosons are the 
$Z, W^{\pm} $ instead of $Z,\gamma $.
We therefore expect for the sneutrinos a smaller 
hadronic branching ratio at low masses, since
in both channels  a massive gauge boson is involved, instead 
of the massless photon. 
In fact in the computation presented in \cite{Steffen:2006hw}
it is apparent that the $Z$ channel is strongly suppressed 
by phase space and remains sub-leading as long as the NLSP mass 
is below $200$ GeV, as in our case.
So we can use the $\epsilon_\text{had} \lesssim 10^{-2}\gev$ value 
given for the $\tilde\tau $ case in \cite{Steffen:2006hw} as
a very stringent 
value also for the sneutrino NLSP\@.
We obtain in this case 
$ \epsilon_\text{had} Y_{\tilde\nu} \leq 4.6 \times 10^{-15}\gev$,
which is well below the hadronic limit (\ref{eq:hadbound}).

If we compare our estimate with the conclusion in 
\cite{Feng:2004mt} that
sneutrinos with masses below $400$ GeV are allowed by
hadronic constraints, we find two differences that 
somewhat change the discussion.
On the one hand, the approximate expression used in~\cite{Feng:2004mt}
to compute the sneutrino abundance always underestimates it 
in our case, probably due to the importance of coannihilations. 
On the other hand, the value of $\epsilon_\text{had} $ used 
there appears overestimated, especially for small sneutrino 
masses, as discussed for the stau case in \cite{Steffen:2006hw}.
The two effects partially compensate each other for small
sneutrino masses.

All in all, we can conclude that the $\gravitino$-$\snu$ scenario
 is essentially unconstrained. The severe electromagnetic 
limits could be marginally relevant and are worth 
a more careful investigation.

\subsection{Constraints from Dark Matter and the CMB}
\label{sec:DMandCMB}

Gravitinos are produced non-thermally via the NLSP decays.
The corresponding energy density has to be smaller than the observed 
cold dark matter density,
\begin{equation}
	\Omega^\text{non-th}_{3/2}h^2 =
	\frac{m_{3/2}}{m_\text{NLSP}} \Omega_\text{NLSP}^\text{th}h^2 \leq
	\Omega_\text{DM} < 0.123 \;.
\end{equation}
For $m_{1/2}=500\gev$ and $\tan\beta=20$, we find
\begin{equation}
\label{eq:nonthY}
	1.8 \cdot 10^{-3} \leq
	\Omega^\text{non-th}_{3/2} h^2 \leq 8.4 \cdot 10^{-3} \;.
\end{equation}
For $\tan\beta=10$ the maximal energy density is slightly smaller.
Generally, $\Omega^\text{non-th}_{3/2}$
is largest in the part of the $\stau$ NLSP region which is closest to the 
$\chi$ LSP region. Here 
we also find the
largest NLSP abundance (cf.\ Fig.~\ref{fig:YStautb20}).
In the $\gravitino$-$\snu$ scenario,
the maximal value for $\Omega^\text{non-th}_{3/2} h^2$ is 
just slightly smaller than in the $\widetilde\tau$ NLSP case.

In addition to bounds from the observed cold dark matter density
there are constraints coming from distortions 
of the cosmic microwave background. 
It is well known that the CMB is very close to a Planckian distribution
with zero chemical potential \cite{Fixsen:1996nj},
\begin{equation}
|\mu| < 9 \cdot 10^{-5} \text{ (at 95\% C.L.)}.
\end{equation} 
Since late electromagnetic
energy release can lead to spectral distortions, this upper limit on
$|\mu|$ can be translated into an upper limit on
$\epsilon_\text{em}$~\cite{Hu:1992dc}.  However, this limit is based
on an approximation which turned out to be reliable only for stau masses
above $500\gev$ in an improved analysis \cite{Lamon:2005jc}.  For
lighter staus, the bounds become weaker.  
As a consequence, they are
less constraining than the BBN bounds in our case \cite{Steffen:2006hw}.

\subsection{Constraints on the Reheating Temperature}

At high temperatures, gravitinos are produced by thermal scatterings.  The
resulting energy density is approximately given by 
\cite{Bolz:2000fu,*Pradler:2006qh}
\begin{equation}
	\Omega_{3/2}^\text{th} h^2 \simeq
	0.27 \left( \frac{T_\text{R}}{10^{10}\gev} \right)
	\left( \frac{100\gev}{m_{3/2}} \right)
	\biggl( \frac{m_{\tilde g}}{1\tev} \biggr)^2 \;,
\end{equation}
where $m_{\tilde g}$ is the running gluino mass evaluated at low energy.
For $m_{1/2}=500\gev$, we have   
$m_{\tilde g}\simeq1150\gev$.
The maximal possible reheating temperature $T_\text{R}$ is obtained for
the heaviest allowed gravitino mass.  In the $\gravitino$-$\stau$
scenario, the gravitino mass is strongly constrained by BBN\@.
The largest allowed value is around $m_{3/2} \simeq 70\gev$.
Using this upper limit and taking as lower bound  
$m_{3/2}=10\gev$, we can calculate an
allowed range for the reheating temperature,
assuming that all the dark matter is made up of
gravitinos.  
Since the non-thermal contribution is negligible (cf.\ Eq.~(\ref{eq:nonthY})),
one obtains from  $\Omega^\text{th}_{3/2} \leq \Omega_\text{DM}$
and  Eq.~(\ref{eq:OmegaDMObs})
\begin{equation}
	3 \cdot 10^8 \gev \lesssim T_\text{R} \lesssim 3 \cdot 10^9 \gev \;.
\end{equation}
This is marginally compatible with the minimal temperature required for thermal
leptogenesis \cite{Buchmuller:2004nz}.
Increasing the unified gaugino mass 
$m_{1/2}$ essentially leads
to a rescaling of the gluino and gravitino masses, which lowers 
the upper bound on $T_\text{R}$.
Therefore a small gaugino mass is needed
for a high reheating temperature.

The $\gravitino$-$\snu$ scenario is less constrained by BBN and therefore
allows for a much heavier gravitino.
The only restriction is that the gravitino be lighter than
the sneutrino,  $m_{3/2}\lesssim 200\gev$. This leads to a larger
allowed range for the reheating temperature,
\begin{equation}
	3 \cdot 10^8 \gev \lesssim T_\text{R} \lesssim 7 \cdot 10^9 \gev \;,
\end{equation}
which is consistent with thermal leptogenesis.

\section{Conclusions}

\begin{figure}
\centering
\includegraphics{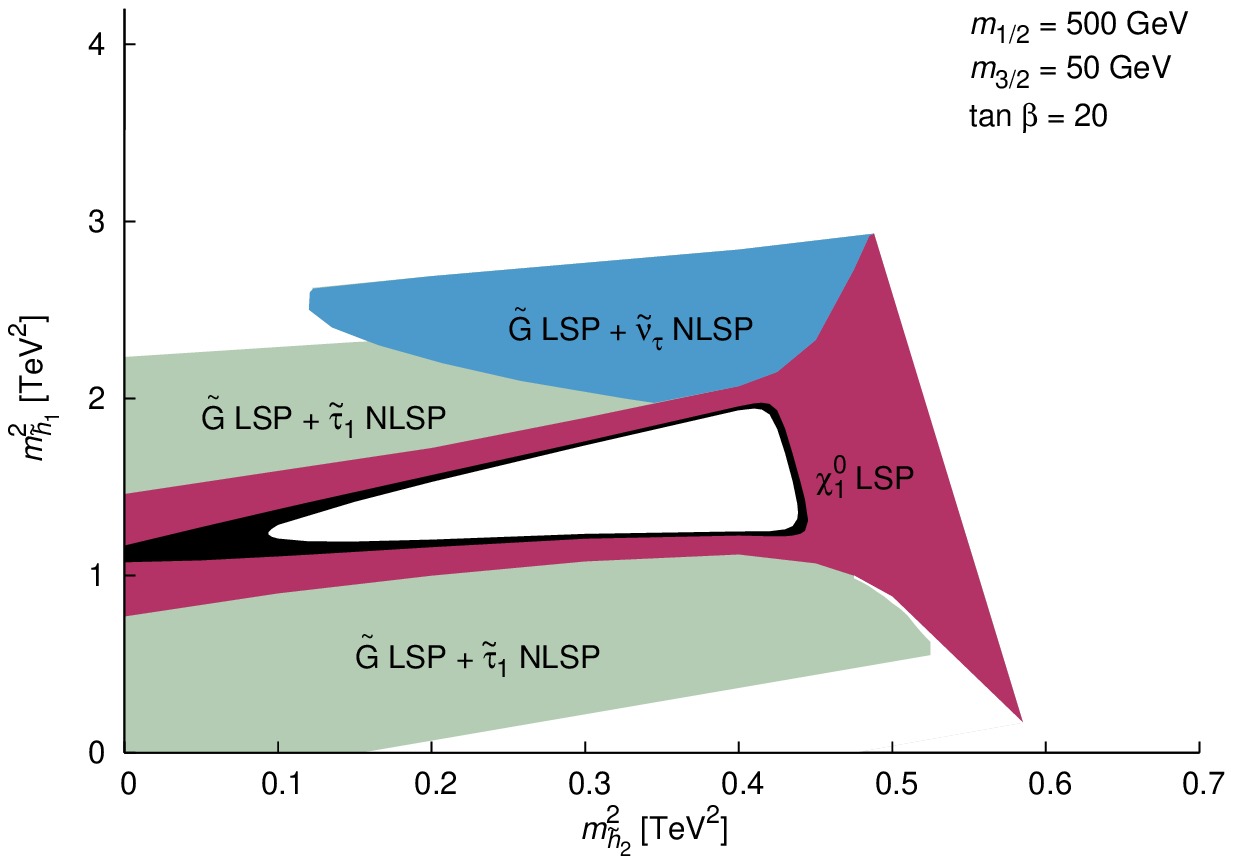}
\includegraphics{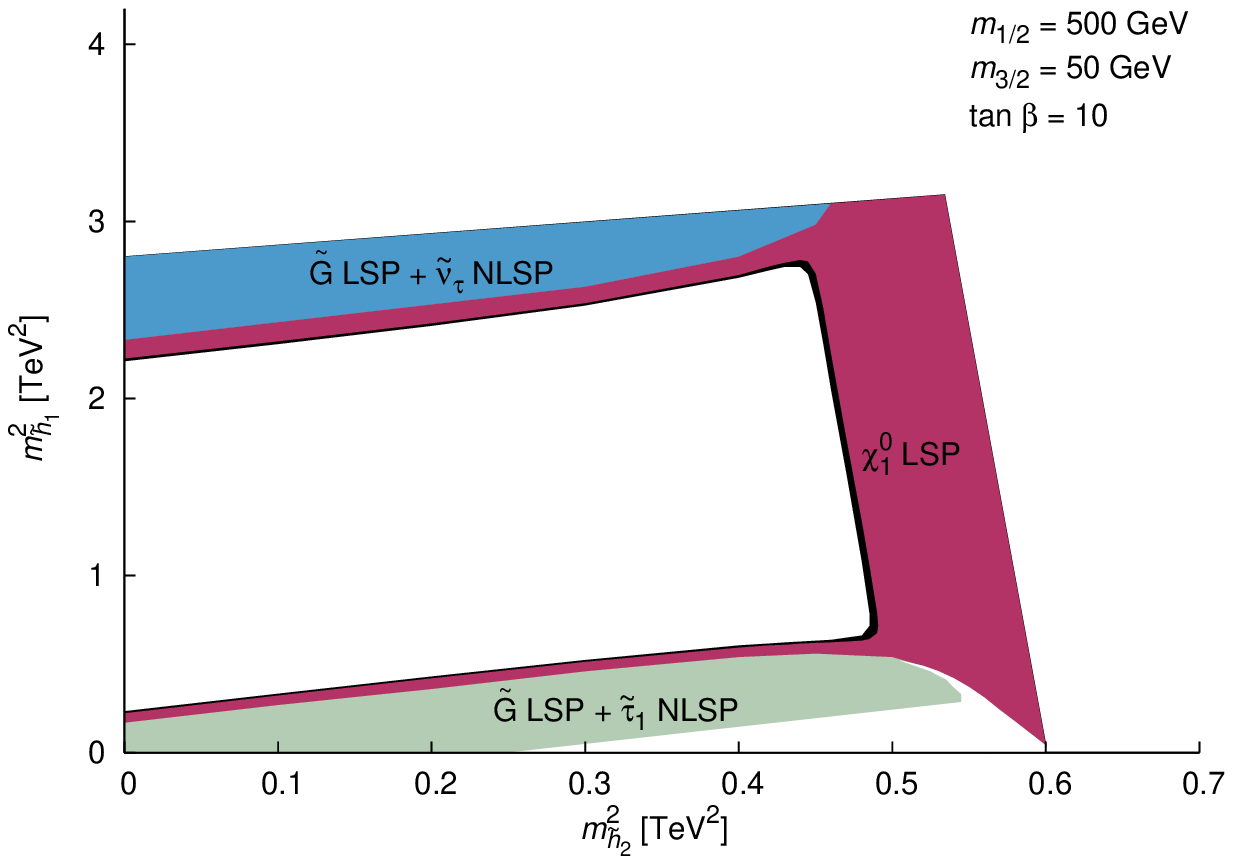}
\caption{Allowed parameter space for the soft Higgs masses 
in gaugino mediation.
In the black and magenta (dark-gray) coloured regions a neutralino is the LSP,
whereas in the green (light-gray) and blue (medium-gray) 
regions the gravitino is the LSP with either 
a stau or a sneutrino being the NLSP\@.
All white areas are excluded by bounds from
the observed dark matter density and the ``conservative''
constraints from primordial nucleosynthesis.
}
\label{fig:finalpara20}
\end{figure}
We have discussed dark matter candidates
in theories with gaugino-mediated 
supersymmetry breaking 
and compact dimensions of order the unification scale.
Varying the boundary conditions for bulk Higgs 
fields at the unification scale, 
at different
points in parameter space, the gravitino, a neutralino or a scalar 
lepton can be the lightest 
or next-to-lightest superparticle. 

We have investigated
constraints from the observed dark matter density and primordial 
nucleosynthesis on the different scenarios. 
The resulting viable dark matter candidates 
in gaugino mediation are summarised
in Fig.~\ref{fig:finalpara20}.
A neutralino LSP as the dominant
component of dark matter is a viable possibility.
Gravitino dark matter with a $\stau$ NLSP is consistent 
for a wide range of parameters only with the ``conservative'' BBN
constraints (cf.\ Eqs.~\eqref{eq:ConservativeBounds}).  The ``severe''
BBN bounds (cf.\ Eqs.~\eqref{eq:SevereBounds}) require either a
gravitino mass close to the lower bound in gaugino mediation
or entropy production after $\stau$ decoupling.
Gravitino dark matter with a $\snu$ NLSP can also be realised and is
essentially unaffected by all constraints.

\section*{Acknowledgements}
We would like to thank Genevieve B\'{e}langer, Nicolao Fornengo, 
Koichi Hamaguchi, Jan Hamann and Frank Steffen
for valuable discussions.
WB thanks the Galileo Galilei Institute for Theoretical Physics
for the hospitality and the INFN for partial support
during the completion of this work.
This work has been supported by the ``Impuls- und Vernetzungsfonds'' of
the Helmholtz Association, contract number VH-NG-006.

\bibliography{GauginoMediation}
\bibliographystyle{ArXivmcite}

\end{document}